\documentclass{article}
\usepackage[utf8]{inputenc}
\usepackage{amsmath}
\usepackage{nicefrac}
\usepackage{amsfonts}
\usepackage{amssymb}
\usepackage{graphicx}
\usepackage{siunitx}
\usepackage{lmodern,bm}
\usepackage{adjustbox}
\usepackage{threeparttable}
\usepackage{placeins}
\usepackage{hyperref}
\usepackage{authblk}

\DeclareSIUnit\pixel{pixel}
\DeclareMathOperator*{\argmin}{\arg\!\min}
\title{Physics-based Reconstruction Methods for Magnetic Resonance Imaging}
\author[1,2]{Xiaoqing Wang}
\author[1,2]{Zhengguo Tan}
\author[1,2]{Nick Scholand}
\author[1]{Volkert Roeloffs}
\author[1,2,3,4]{Martin Uecker}

\affil[1]{Institute for Diagnostic and Interventional Radiology, University Medical Center Göttingen, Göttingen, Germany}
\affil[2]{German Centre for Cardiovascular Research (DZHK), Göttingen, Germany}
\affil[3]{Cluster of Excellence “Multiscale Bioimaging: from Molecular Machines to Networks of Excitable Cells” (MBExC), Göttingen, Germany}
\affil[4]{Campus Institute Data Science (CIDAS), Göttingen, Germany}

\begin{document}
\maketitle

\begin{abstract}
Conventional Magnetic Resonance Imaging (MRI) is hampered by long
scan times and only qualitative image contrasts that prohibit a direct 
comparison between different systems. To address these limitations,
model-based reconstructions explicitly model the
physical laws that govern the MRI signal generation. By formulating image
reconstruction as an inverse problem, quantitative maps of the underlying physical
parameters can then be extracted directly from efficiently acquired k-space signals without intermediate image reconstruction -- addressing both shortcomings of conventional
MRI at the same time. This review will discuss basic concepts of model-based
reconstructions and report about our experience in developing
several model-based methods over the last decade using
selected examples that are provided complete with data
and code.
\end{abstract}
\section{Introduction}

First physics-based reconstruction methods for parametric mapping
appeared in the literature more than a decade ago
\cite{Graff__2006,Olafsson_IEEETrans.Med.Imag._2008,Block_IEEETrans.Med.Imaging_2009}
and constitute now a major research area in the field of Magnetic Resonance Imaging (MRI)
\cite{Fessler_IEEESignalProcess.Mag._2010,
	Sumpf_J.Magn.Reson.Imaging_2011,
	tran_Magn.Reson.Med.2013,
	Zhao_IEEETrans.Med.Imag._2014,
	Tran-Gia_PLOS_ONE_2015,
	Roeloffs_Int.J.Imag.Syst.Tech._2016,
	Ben-Eliezer_Magn.Reson.Med._2016,
	Tan_Magn.Reson.Med._2017,
	Sbrizzi_arXiv_2017,
	Wang_Magn.Reson.Med._2018}.
Model-based reconstruction is based on modelling the physics of the MRI signal
and has been used, for example, to estimate T1~\cite{Doneva_Magn.Reson.Med._2010,tran_Magn.Reson.Med.2013, Roeloffs_Int.J.Imag.Syst.Tech._2016, Tran-Gia_Magn.Reson.Imaging_2016, Wang_Magn.Reson.Med._2018,Maier_Magn.Reson.Med._2018}, T2 relaxation~\cite{Block_IEEETrans.Med.Imaging_2009,Sumpf_J.Magn.Reson.Imaging_2011,Ben-Eliezer_Magn.Reson.Med._2016,Hilbert_J.Magn.Reson.Imag._2018}, for T2$^{\star}$ estimation
and water-fat separation~\cite{Doneva_Magn.Reson.Med._WF_2010,Wiens_Magn.Reson.Med._2014, zimmermann2017accelerated, Benkert_Magn.Reson.Med._2017,Schneider_Magn.Reson.Med._2020}, as well as for quantification of flow~\cite{Tan_Magn.Reson.Med._2017} and diffusion~\cite{Welsh_Magn.Reson.Med._2013,knoll_NMR_Biomed.2015}.
Quantitative maps of the underlying physical parameters can then be extracted directly
from the measurement data without intermediate image reconstruction. 
This direct reconstruction has two major advantages: First, the full signal is described by a 
model based on few parameter maps only and intermediate image reconstruction is waived. This renders
model-based techniques much more efficient in exploiting the available information than 
conventional two-step methods. Second, as a specific signal behaviour is no longer required
for image reconstruction, MRI sequences can now be designed that have an optimal sensitivity to the parameters of interest.
Once the underlying physical parameters are estimated, arbitrary contrast-weighted
images can be generated synthetically by evaluating the signal model for a specific
sequence and acquisition parameters.

In this work, we discuss our experience using model-based reconstruction methods
with different radial MRI sequences showing a variety of examples ranging from
T1 and T1 mapping and banding-free bSSFP imaging in the brain over
flow quantification in the aorta to water-fat separation and $T_2^\star$ mapping in the liver.
All provided examples come with data and code and can be 
reconstructed using the BART toolbox \cite{Uecker__2015}.

\section{MRI Signal}

In typical MRI experiments, the proton spins are polarized by
bringing them into a strong external field. The spins then start
to precess with a characteristic Larmor frequency and can be manipulated
using additional on-resonant radio-frequency pulses and 
further gradient fields. The dynamical behaviour of the
magnetization is described by the Bloch-Torrey equations that
describe the physics of magnetic resonance including effects from
relaxation, flow and diffusion. As a fully computer-controlled
imaging method, MRI is extremely flexible and the underlying physics
enables access to a variety of tissue and imaging system specific parameters such as 
relaxation constants, flow velocities, diffusion, temperature,
magnetic fields, etc.

The measured MRI signal corresponds to the
complex-valued transversal magnetization $M$ 
which is obtained by quadrature demodulation from the voltage induced
in the receive coils.
In a multi-coil experiment this signal is proportional to the
transversal magnetization weighted by the sensitivity of each receive coil:
\begin{equation}
\label{MR_signal}
	y_{j}(t) = \int c_{j}(\vec{r}) \, M(x, B, t, \vec{r}) \, d\vec{r}
\end{equation}
Here, $c_j$ is the complex-valued
sensitivity of the $j$th coil and $M$ the complex-valued
transversal magnetization at time $t$ and position $\vec{r}$.
The magnetization depends on some physical parameters $x$
and the externally controlled magnetic fields $B(t, \vec{r})$,
i.e. gradient fields and radio-frequency pulses,
and can be obtained by solving the Bloch-Torrey equations
(or, if motion of spins can be neglected, by solving the Bloch
equations at each point).

\begin{figure}
	\begin{center}
		\includegraphics[width=\textwidth]{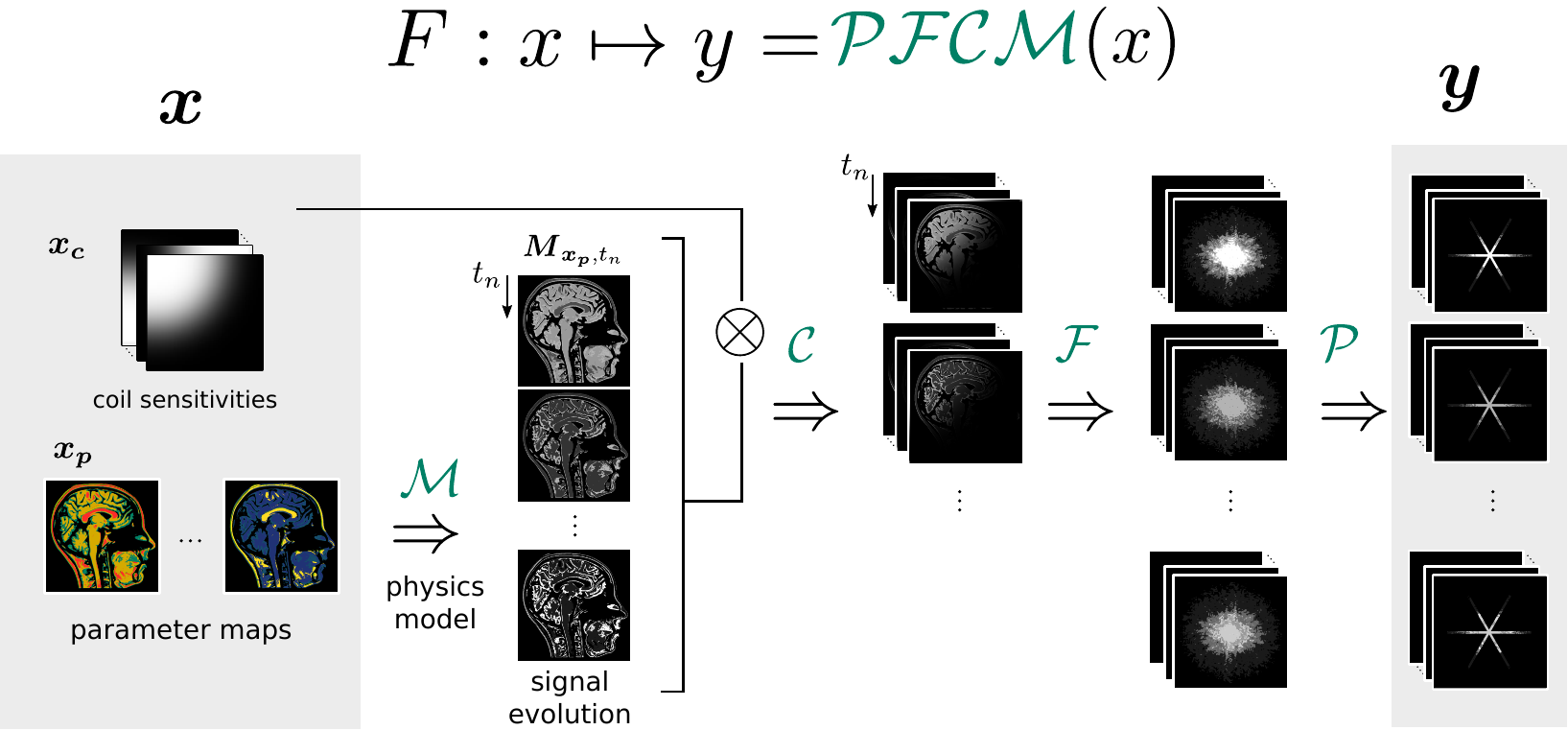}
	\end{center}
	\caption{The forward operator $F$
		can be formally factorized into operator $\mathcal{M}$ that describes the 
		spin physics, the multiplication with the coil sensitivities $\mathcal{C}$, 
		the (non-uniform) Fourier transform $\mathcal{F}$, and a sampling
		operator $\mathcal{P}$.}
	\label{fig:forward}
\end{figure}

\begin{table}
	\renewcommand{\arraystretch}{1.4}
	\begin{center}
		\begin{tabular}[h]{c|c|c}
			parameter & sequence type & signal\\
			\hline
			relaxation rate $R_1$ &	inversion recovery &	$a - (1 + a)\cdot e^{-t_{n}\nicefrac{R_{1}}{a}}$ \\
			relaxation rate $R_2$ & spin-echo &	$e^{-R_{2} t_{n} }$ \\
			relaxation rate $R_2^\star$ & gradient-echo & $e^{ - {R_2^*}  t_n}$ \\
			field $B_0$ & gradient-echo & $e^{i 2\pi \cdot f_{B_0} t_n}$ \\
			chemical shift & gradient-echo & $\sum_p e^{i 2\pi f_p t_n}$ \\
			flow velocity $\vec v$ & bipolar gradient & $e^{i \vec v \cdot \vec V_n}$ \\
			diffusion tensor $D$ & bipolar gradient & $e^{-\vec b_{n}^T D b_n}$ 
		\end{tabular}
		\caption{Basic analytical signal models for physical parameter
			dependencies in common MRI sequences.}
		\label{sigmodels}
	\end{center}
\end{table}

While Equation (\ref{MR_signal}) can be exploited directly for model-based reconstruction \cite{Sbrizzi_MagnResonImaging_2018}, many other model-based 
methods use some simplifying approximations to reduce the computational complexity.
Often, segmentation in time is used by assuming that the
magnetization is constant around certain time points, 
e.g. around echo times $\text{TE}_n$ with $n \in 1, \dots, N$. $N$ is the number of echos.
The effect of magnetic field gradients can then be separated
out into a phase term which is defined by the k-space
trajectory $\vec k(t)$.
This separation often allows the use of simplified models for the
magnetization and, more importantly, the use of fast (non-uniform)
Fourier transform algorithms for the gradient-encoding term.
We derive the following - still very generic - model:
\begin{align}
	y_{n,j}(t) & = \int e^{i 2 \pi \vec{r}\cdot \vec{k}(t)} c_{j}(\vec{r}) M(x, t_n, \vec{r}) d\vec{r} 
\end{align}
Based on this model, we define a nonlinear forward operator $F: x \mapsto y$ that
maps the unknown parameters $x$ to the acquired data $y$.
This operator can be formally decomposed into $F = \mathcal{P F C M}$, where $\mathcal{P}$ is the
sampling operator, $\mathcal{F}$ is the Fourier transform, $\mathcal{C}$ the
multiplication with the coil sensitivities, and $\mathcal{M}$ the signal model (see Fig.~\ref{fig:forward}).

The defining feature of physics-modelling reconstruction is the addition
of a signal model $\mathcal{M}$ into the forward operator $F$. The specific signal model depends on
the applied sequence protocol and specifies which tissue and/or hardware
characteristics can be estimated. Often, an analytical model can be
derived from the Bloch equations using hard-pulse approximations.
For many typical MRI sequences important parameter dependencies are
exponentials. Table~\ref{sigmodels} lists some of these
analytical signals models.
If the applied sequence protocol does not lend itself to an analytical signal expression, the Bloch equations need to be integrated as signal model directly. This integration becomes challenging for iterative reconstructions, because of the estimation of the signals derivatives. Current techniques exploit finite difference methods \cite{Ben-Eliezer_Magn.Reson.Med._2016, Sbrizzi_MagnResonImaging_2018} or sensitivity analysis of the Bloch equations \cite{Scholand__2020a}.

\vspace{0.5cm}
\begin{threeparttable}
	\renewcommand{\arraystretch}{1.4}
	\centering
	\begin{adjustbox}{width=1\textwidth}
		\begin{tabular}[h]{c|c|c|c|c|c|c|c|c|c}
			Sequence & Flip Angle		& TR/TE/$\Delta$TE	& Bandwidth 	& Matrix  & Spokes  & TA 	& FOV   &  Slice	\\
			& \si{\degree} 		& ms 			& Hz/px &	&	&  s	& mm   &  mm	\\
			\hline
			IR-FLASH & 	6	& 4.10 / 2.58 		& 630 	& 256 $\times$ 256   & 1020 	& 4		& 192 	& 5 \\
			ME-SE	 & 90/180	& 2500 / 9.9 / 9.9  	& 390 	& 256 $\times$ 256	& 25 $\times$ 16	& 80			& 192	& 3 \\
			ME-FLASH & 5		& 10.60 / 1.37 / 1.34	& 960	& 200 $\times$ 200	& 33 $\times$ 7	& 0.35\tnote{ 1}	& 320	& 5 \\
			PC-FLASH & 10		& 4.46 / 2.96 		& 1250  & 210 $\times$ 210	& 2 $\times$ 7	& 15		& 320 	& 5 \\
			fmSSFP\tnote{2} & 15 & 4.5  / 2.25 		&  840  & 192 $\times$ 192   & 4 $\times$ 101 $\times$ 40  	& 137	& 192	& 1
		\end{tabular}
	\end{adjustbox}
	
	\begin{tablenotes}
		\item[1] acquisition time per frame, because the presented example is based on \\
		dynamic acquisition
		\item[2] 3D Stack-Of-Stars sequence
		with 40 partitions (1000 prep scans), while all \\
		other acquisition protocols are 2D
	\end{tablenotes}
	
	\caption{Detailed parameters of MR sequences capable of mapping physical parameters listed in Table \ref{sigmodels}.}
	\label{seqparm}
	
\end{threeparttable}
\vspace{0.5cm}

The MRI data used in this work was acquired on a Siemens Skyra 3T scanner (Siemens Healthcare GmbH, Erlangen, Germany)
from four volunteers (21-35 years, two females) without known illness after obtaining written informed consent and with
approval of the local ethics committee. Acquisition parameters can be found in Table~\ref{seqparm}.

\section{Nonlinear Reconstruction}

Using a nonlinear forward operator $F: x \mapsto y$ that maps the
unknown parameters $x$ to the acquired data $y$, we can formulate the image
reconstruction as a nonlinear optimization problem:
\begin{align}
\label{moba_eq}
    \hat{x} = \argmin_{x} \|F(x) - y\|_{2}^{2} + \sum_{i}\lambda_{i}R_{i}(x)
\end{align}
Data fidelity is ensured by $\|F(x) - y\|_{2}^{2}$ and regularization 
terms $R_i$ can be added to introduce prior knowledge with $\lambda_{i}$ the corresponding 
regularization parameters. This framework
is very general, combining parallel imaging, compressed sensing, and
model-based reconstruction in a unified reconstruction.
Often the coil sensitivities are estimated before, but they could
also be included as unknowns in $x$. Paired with suitable
sampling schemes, this yields fully calibrationless methods
that do not require additional calibration scans \cite{Wang_Magn.Reson.Med._2018,Wang_J.Cardiovasc.Magn.Reson._2019,Wang_Magn.Reson.Med._2020,Tan_Magn.Reson.Med._2017,Tan_NMRBiomed._2017,Tan_Magn.Reson.Med._2019}. Moreover, model-based reconstructions allow a direct 
application of sparsity-promoting regularizations to the physical parameters for performance
improvement \cite{Block_IEEETrans.Med.Imaging_2009,Zhao_IEEETrans.Med.Imag._2014,knoll_NMR_Biomed.2015, Wang_Magn.Reson.Med._2018}. 

However, the high non-convexity of model-based reconstruction makes this method sensible to the
initial guess and relative scaling of the derivatives of each parameter map. These 
issues can often be addressed with a reasonable initial guess
and a proper preconditioning. Algorithms to solve the nonlinear inverse problems include 
gradient descent, the variable projection methods \cite{Golub_InverseProb_2003,Zhao_IEEETrans.Med.Imag._2014}, the method of nonlinear conjugate gradient \cite{Hager_SIAMJ.Optim._2005} and
Newton-type methods \cite{Bakushinsky__2005}. Particularly for the examples presented in this paper, we solve Equation (\ref{moba_eq}) via an iteratively regularized Gauss-Newton method (IRGNM) \cite{Bakushinsky__2005} where the nonlinear
problem in Equation (\ref{moba_eq}) is linearized in each Gauss-Newton step, i.e.
\begin{align}
\label{linear_eq}
\hat{x}_{n+1} = \argmin_{x} \|DF(x_{n})(x-x_{n}) + F(x_{n}) - y\|_{2}^{2} + \sum_{i}\lambda_{i}R_{i}(x)
\end{align}
with $DF(x_{n})$ the Jacobian matrix of $F$ at the point $x_{n}$. The regularized linear subproblem can be further solved by conjugate gradients, FISTA \cite{Beck_SIAMJ.Img.Sci._2009} or
ADMM\cite{Boyd_Found.TrendsMach.Learn._2011}.

\subsection{T1 and T2 Mapping}

\begin{figure}
	\begin{center}
		\includegraphics[width=\textwidth]{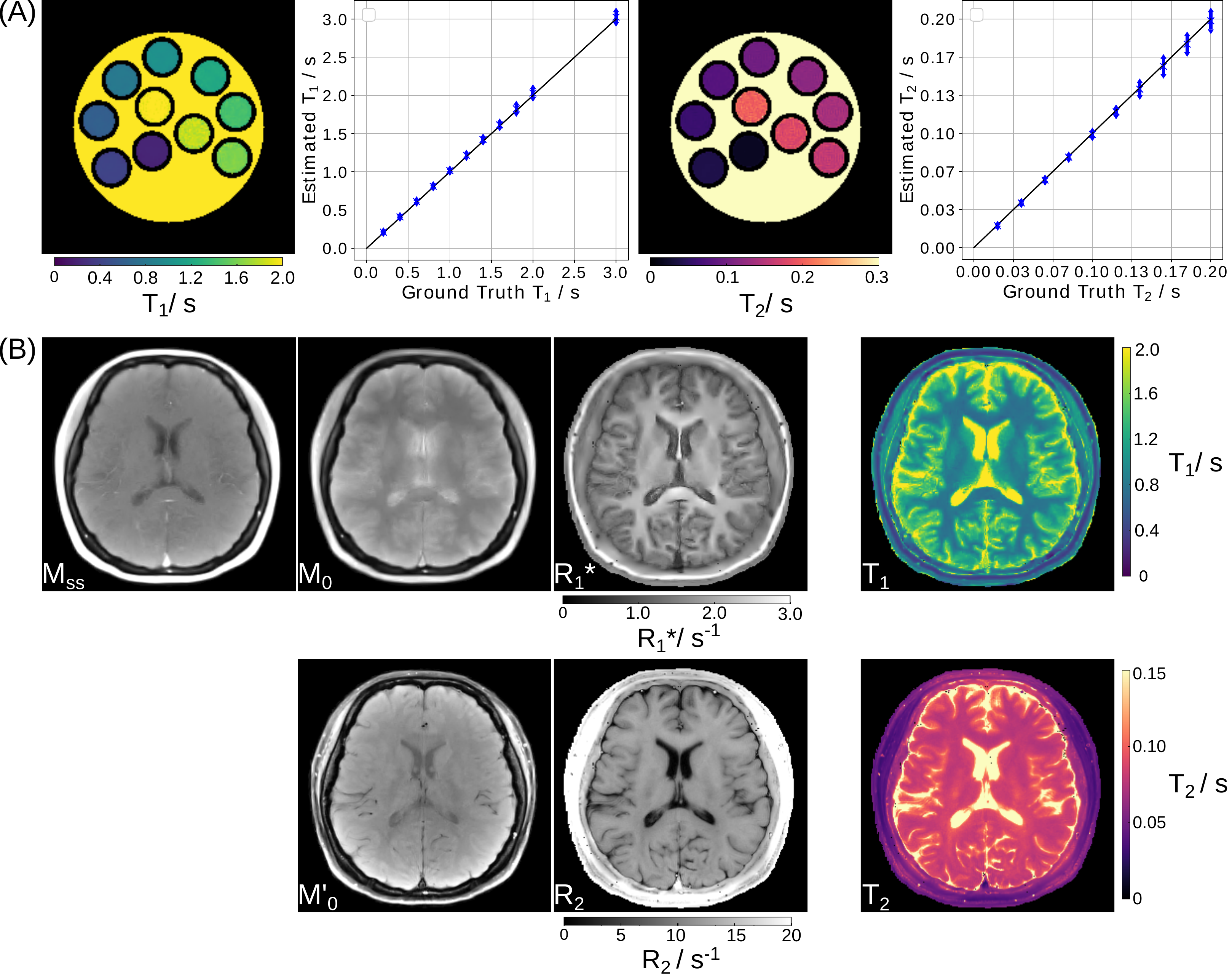}
	\end{center}
	\caption{
		\textbf{(A)}. (Leftmost) Model-based reconstructed T1 map and (left middle) the 
		ROI-analysed quantitative T1 values for the numerical phantom using the
		single-shot IR radial FLASH sequence. (Right middle and rightmost) Similar results for T2 mapping using the
		multi-echo spin-echo sequence.
		\textbf{(B)}. (Top) The reconstructed parameter maps $(M_{ss}, M_{0}, R_{1}^{*})^{T}$ for the T1 model
		and (bottom) $(M_{0}, R_{2})^{T}$ for the T2 model with the corresponding T1 / T2 maps 
		in the rightmost column.}
	\label{fig:nlt1t2}
\end{figure}

T1 mapping can be accomplished using a inversion-recovery (IR) FLASH sequence:
Following a inversion pulse, data is continuously
acquired using the FLASH readout.
The magnetization signal $M(t)$ for 
IR-FLASH reads \cite{Look_Rev.Sci.Instrum._1970, Deichmann_J.Magn.Reson._1992}
\begin{align}
	M_{t_{k}} (\vec{r}) = M_{ss}(\vec{r}) - \big(M_{ss}(\vec{r}) + M_{0}(\vec{r})\big)\cdot e^{-t_{k}\cdot R_{1}^{*}(\vec{r})}	\label{signal_T1}
\end{align}
with $M_{ss}$ the steady-state magnetization, $M_{0}$ the equilibrium
magnetization, and $R_{1}^{*}$ the effective relaxation rate, i.e. $R_{1}^{*} = 1/T_{1}^{*}$.  $t_{k}$ is the
inversion time defined as the center of each acquisition window. The acquisition window is determined by the product of repetition time and the number of readouts formulating one k-space (binning) after inversion. Although model-based reconstructions don't require any binning, this process is helpful for reducing the computation demand while still keeping the $T_{1}$ accuracy \cite{Wang_Magn.Reson.Med._2018}. After estimation of $M_{ss}, M_{0}$ and $R_{1}^{*}$, $T_{1}$ then can be calculated afterwards by: $T_{1} = M_{0} /(M_{ss}\cdot R_{1}^{*})$. Data for T1 mapping is
acquired using a single-shot IR radial FLASH (4 seconds) sequence with a tiny golden angle ($\approx
23.36^{\circ}$) between successive spokes. 

The multi-echo spin echo (ME-SE) sequence can be employed for T2 mapping. The magnetization signal $M(t)$ for a multi-echo spin echo sequence at echo time $t_{k}$ follows
an exponential decay $M_{t_{k}} (\vec{r}) = M'_{0}\cdot e^{-t_{k}\cdot R_{2}(\vec{r})}$
with $M'_{0}$ the spin density map, $R_{2} = 1/T_{2}$ the transverse relaxation rate.
This simple exponential model does not take stimulated echoes into account, but a more
complicated analytical model exists for this case \cite{Sumpf_IEEETrans.Med.Imaging_2014,Petrovic_J.Mange.reson_2019}. Data for T2 mapping is
obtained with 25 excitations and 16 echoes per excitation using a radial golden-ratio ($\approx
111.25^{\circ}$) sampling strategy.

Quantitative parameter maps for both acquisitions are
estimated using the nonlinear model-based reconstruction. In other words, the estimation
of parameter maps $(M_{ss}, M_{0}, R_{1}^{*})^{T}$ or parameter maps $(M'_{0},R_{2})^{T}$, respectively,
and coil sensitivity maps $(c_{1}, \ldots, c_{N})^{T}$ is formulated as a nonlinear inverse problem with
a joint $\ell_{1}$-Wavelet regularization applied to the parameter maps and the
Sobolev norm \cite{Uecker_Magn.Reson.Med._2008} to the coil sensitivity maps. This nonlinear inverse problem is
then solved by the IRGNM-FISTA algorithm \cite{Wang_Magn.Reson.Med._2018}.
After estimation of the parameters $T_{1}$ and $T_{2}$ maps can be calculated.
Note that the $M_{0}$ and $M'_{0}$ absorb physical effects which are not
explicitely modelled and identical over all inversion or echo times.

To evaluate the quantitative accuracy of the model-based methods, numerical
phantoms with different T1 relaxation times (ranging from 200 ms to 2000 ms
with a step size of 200 ms for each tube, and 3000 ms for the background), T2
relaxation times (ranging from 20 ms to 200 ms with a step size of 20 ms for
each tube, and 1000 ms for the background) were simulated, respectively. To
avoid an inverse crime \cite{Colton__2013}, the $k$-space data was derived from the analytical Fourier representation of an
ellipse assuming an array of eight circular receiver coils surrounding the
phantom without overlap. Complex white Gaussian noise with a moderate standard deviation was added to the simulated $k$-space data.

Fig.~\ref{fig:nlt1t2} (A) presents the estimated T1, T2 maps and the corresponding 
ROI-analyzed quantitative values for the numerical 
phantom using model-based reconstructions. Good quantitative accuracy is confirmed
for both model-based T1 and T2 mapping methods. Fig.~\ref{fig:nlt1t2} (B) demonstrates
model-based reconstructed three and two physical parameter maps, the
corresponding T1 and T2 maps for the retrospective T1 and T2 models on human
brain studies.  Further, synthetic images were computed for all inversion/echo
times and the image series was then converted into movies showing the contrast
changes in Supplementary Videos 1 and 2. For the data presented here, 
model-based T1 and T2 reconstruction 
took around 6 and 3 minutes on a GPU (Tesla V100 SXM2, NVIDIA, Santa Clara, CA), respectively.

\subsection{Water/Fat Separation and $R_2^*$ Mapping}

\begin{figure}
	\begin{center}
		\includegraphics[width=\textwidth]{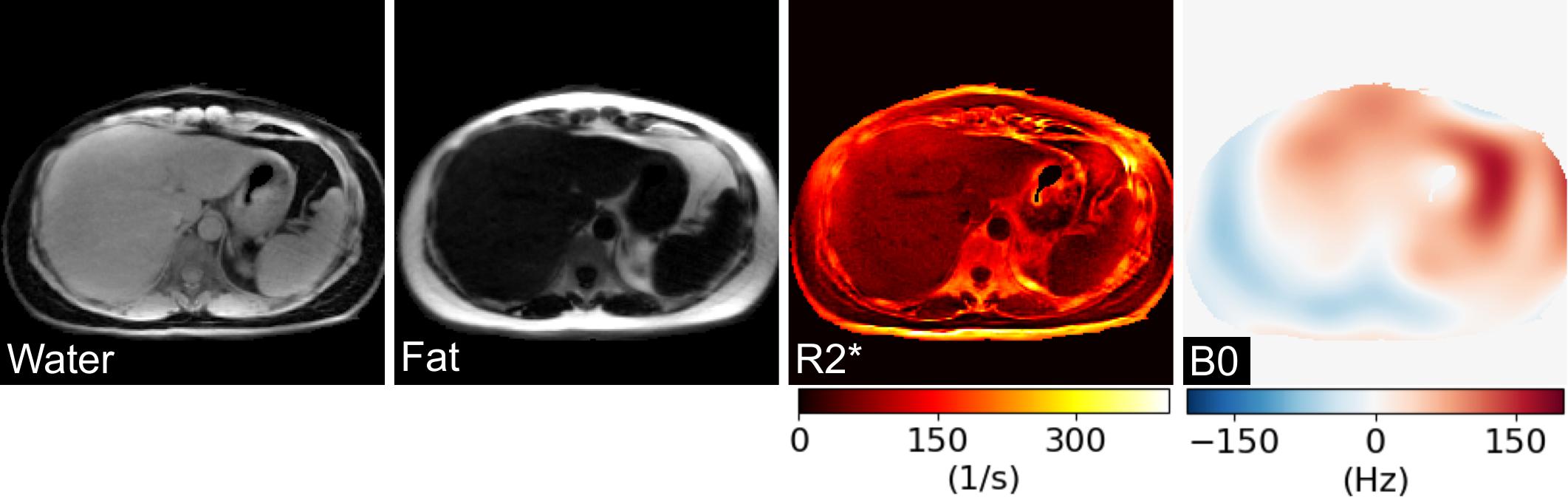}
	\end{center}
	\caption{Real-time liver images acquired during free-breathing using a radial multi-echo (ME) FLASH sequence.
		Model-based reconstruction directly and jointly estimates separated water and fat images, as well as $R_2^{\star}$ and $B_0$ field maps.}
	\label{fig:wf}
\end{figure}

Quantitative $T_2^*$ mapping can be achieved via multi-echo gradient-echo sampling.
With prolonged echo-train readout, the acquired multi-echo signal is
\begin{equation}
    \label{signal_r2s}
    M_{n} = \rho \cdot e^{ - {R_2^*} \cdot \text{TE}_n} \cdot e^{i 2\pi \cdot f_{B_0} \cdot \text{TE}_n}
\end{equation}
where $R_2^*$ is the inverse of $T_2^*$ and $f_{B_0}$ is the $B_0$ field
inhomogeneity. $\text{TE}_n$ denotes the $n$th echo time. On the
other hand, when the imaging voxel contains distinct protons resonating at
different frequencies, the magnetization $\rho$ can be split into multiple
compartments. For instance, chemical shift between water and fat induces phase
modulation, therefore, 
\begin{equation}
    \label{signal_wfr2s}
    M_{n} = (\text{W} + \text{F} \cdot \sum_p e^{i 2\pi f_p \cdot \text{TE}_n}) \cdot e^{ - {R_2^*} \cdot \text{TE}_n} \cdot e^{i 2\pi \cdot f_{B_0} \cdot \text{TE}_n}
\end{equation}
where W and F are the water and fat magnetization, respectively. $f_p$ is the
$p$th fat-spectrum peak frequency. In practice, usually the 6-peak fat spectrum
\cite{Hu_Magn.Reson.Med._2012} is used. In the model-based reconstruction formulation, 
the unknowns contain $\text{W}$, $\text{F}$, $R_2^*$, and $f_{B_0}$, as well as a set of 
coil sensitivity maps from the parallel imaging model.

Here, a multi-echo (ME) radial FLASH sequence \cite{Tan_Magn.Reson.Med._2019} was used
to acquire liver data during free breathing. 
The model-based reconstruction was initialized by the estimate from 
model-based 3-point water/fat separation \cite{Tan_Magn.Reson.Med._2019}, 
while $R_2^*$ and coil sensitivity maps were initialized with 0. 
Afterwards, joint estimation of all unknowns in Equation (\ref{signal_wfr2s}) 
including coil sensitivity maps was achieved via IRGNM with ADMM.
The Sobolev-norm weight \cite{Uecker_Magn.Reson.Med._2008} 
was applied to the $B_0$ field inhomogeneity and coil sensitivity maps. 
Joint $\ell1$-Wavelet regularization was applied to other parameter maps. 
As shown in Figure~\ref{fig:wf} and Supplementary Video 3, high-quality
respiratory-resolved water/fat separation as well as $R_2^*$ and $f_{B_0}$ maps
can be achieved even with undersampled multi-echo radial acquisition (33 spokes
per echo and 7 echoes in total).

\subsection{Phase-Contrast Velocity Mapping}

In phase-contrast flow MRI, velocity-encoding gradients 
are used to encode flow-induced phases.
Due to the complexity of MR signal, 
a reference measurement without flow-encoding gradients is
required such that the phase difference between these two 
measurements excludes the background phase. Therefore, 
the phase-contrast flow MRI signal can be modeled as
\begin{equation}
    \label{signal_phasediff}
    M_{k} = \rho \cdot e^{i \vec v \cdot \vec V_k} \; .
\end{equation}
$\vec v$ is the velocity and $V_k$ is the velocity-encoding for the $k$th measurement. 
For through-plane velocity mapping, $V_0 = 0$ for the reference and $V_1 = \pi / \text{VENC}$
the velocity-encoded measurement, respectively.
VENC is the maximum measurable velocity.
$\rho$ is the shared anatomical image between the two measurements.

\begin{figure}
	\begin{center}
		\includegraphics[width=\textwidth]{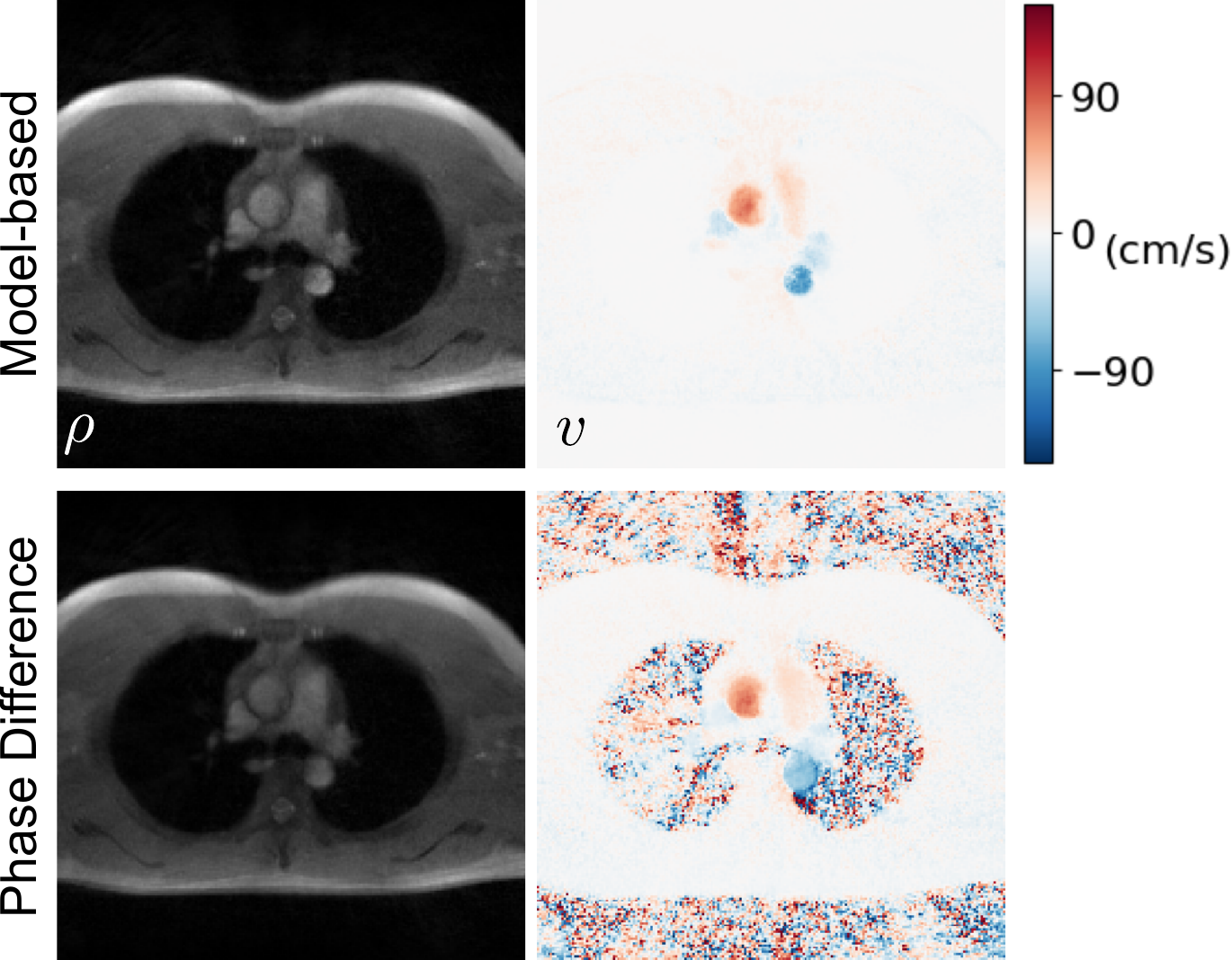}
	\end{center}
	\caption{Comparison between (top) the model-based reconstruction and (bottom) the conventional phase-difference reconstruction. A section crossing the ascending and descending aorta was selected as the imaging slice. Displayed images are (left) anatomical magnitude image and (right) phase-contrast velocity map at systole. With direct phase-difference regularization, the model-based reconstruction largely reduces random background phase noise in the velocity map.}
	\label{fig:flow}
\end{figure}

As an example, flow MRI sequence with radial sampling and through-plane
velocity-encoding gradient was used to measure aortic blood flow velocities. 
As shown in Fig.~\ref{fig:flow}, 
with direct regularization on the phase-difference map, the proposed model-based
reconstruction \cite{Tan_Magn.Reson.Med._2017,Tan_NMRBiomed._2017} is able to 
largely remove background random phase noise. 
Supplementary Video 4 displays the dynamic velocity maps of the whole 15-second scan.

\section{Linear Subspace Reconstruction}

In contrast to nonlinear models, in linear subspace methods the
signal curves $t \mapsto M(x, t, \vec r)$ are 
approximated using a linear combination of basis functions
\cite{petzschner_Magn.Reson.Med._2011,huang_Magn.Reson.Med.2012,Mani_Magn.Reson.Med._2015,Tamir_Magn.Reson.Med._2017,
Asslaender_Magn.Reson.Med._2018,Roeloffs_Magn.Reson.Med._2018,Pflister_Magn.Reson.Med._2019,Roeloffs_IEEETrans.Med.Imag._2020,Dong_Magn.Reson.Med._2020}.
, i.e.
\begin{align}
	M(x, t, \vec r) \approx \sum_s a_s(\vec r) B_{s}(t)~.
\end{align}
The linear basis functions $B_{s}(t)$ can be generated by simulating a
set of representative signal curves for a range of parameters and performing
a singular value decomposition to obtain a good representation. 

With known coil sensitivities, this leads to a linear inverse problem
for the subspace coefficients:
\begin{align}
	\hat{a} = \argmin_{a} \|\mathcal{P} \mathcal{F} \mathcal{C} B a - y\|_{2}^{2} + \sum_{i}\lambda_{i}R_{i}(a)
\end{align}

After reconstruction of the subspace coefficients $a_s$, the
parameters $x$ need to be estimated in a separate step. This can
be achieved by predicting complete magnetization maps for
all time points and fitting a nonlinear signal model.
This can be done point-wise, so is much easier than doing
a full nonlinear reconstruction. Still, for multi-parametric
mapping efficient techniques to map between coefficients
and parameters are required \cite{Roeloffs_IEEETrans.Med.Imag._2020}.

Linear subspace methods have several advantages.
Linear subspace models lead to linear inverse problems which do
not have local minima. Due to their linearity, they also inherently
avoid model violations stemming from partial volume effects.
Because the matrix multiplication with the basis commutes
with other operations that are identical at each time point, 
is is possible to combine the basis with the sampling
operator. The reconstruction then admits a computationally 
advantageous formulation that allows computation to performed
entirely in the subspace \cite{Mani_Magn.Reson.Med._2015,Tamir_Magn.Reson.Med._2017}.

\subsection{T1 Mapping}

\begin{figure}
	\begin{center}
		\includegraphics[width=\textwidth]{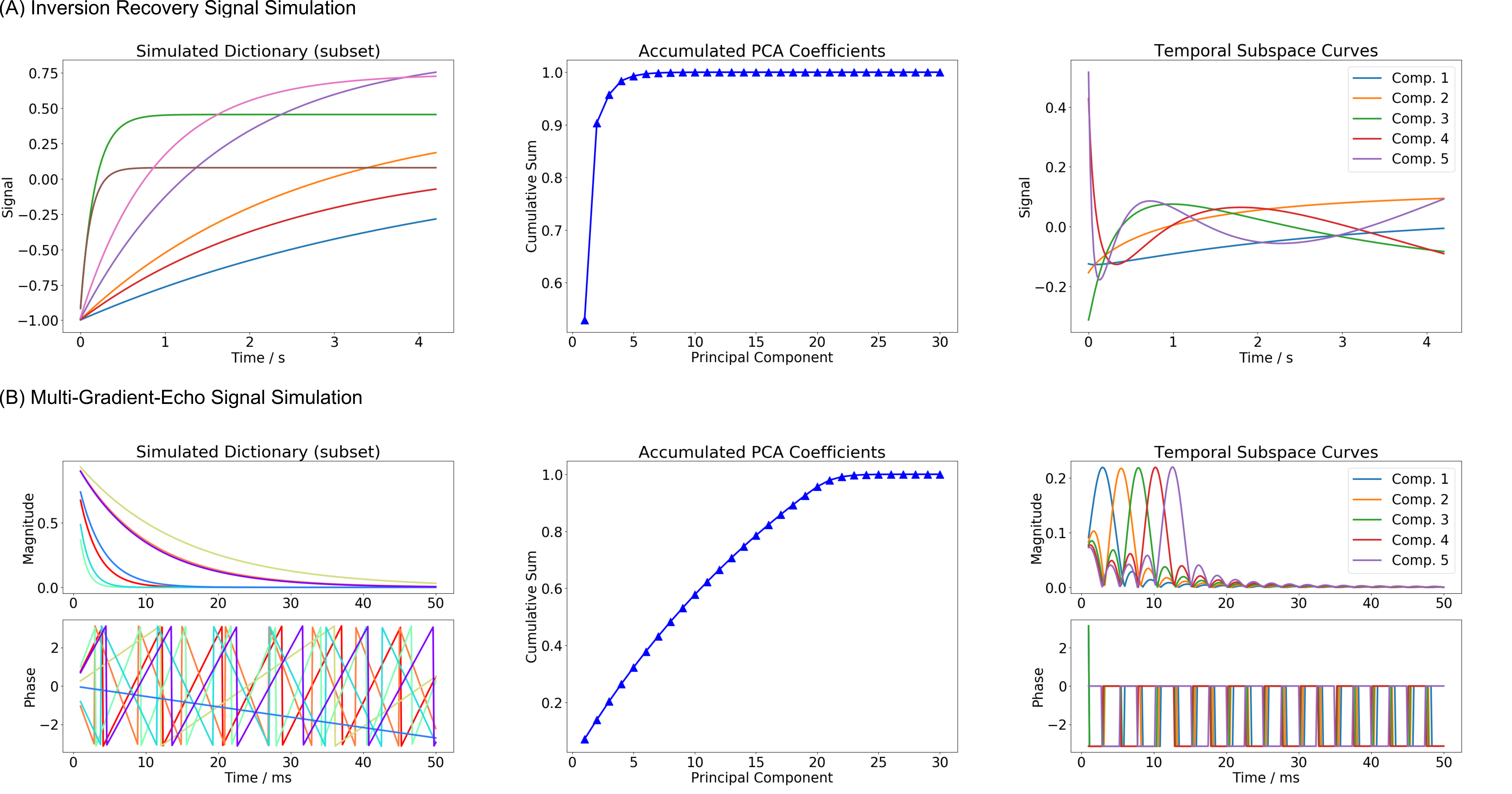}
	\end{center}
	\caption{Demonstration of subspace-based methods for (A) single-shot inversion-recovery and (B) multi-gradient-echo signal, respectively. (Left) Simulated (top) T1
		relaxation and (bottom) $T_2^*$ relaxation and off-resonance phase modulation curves. (Center) Plot of the first 30 principle components.
		(Right) The temporal subspace curves that can be linearly combined to form (top) T1 relaxations and (bottom) multi-gradient-echo relaxations.}
	\label{linplot}
\end{figure}

Alternatively, T1 maps were also reconstructed using the subspace method.
Similar to \cite{Pflister_Magn.Reson.Med._2019}, the T1 dictionary was
constructed using 1000 different $1/R_{1}^{*}$ values linearly range from
5–5000 ms, combining with 100 $M_{ss}$ values from $0.01\cdot M_{0}$ to $
M_{0}$. This results in 100,000 exponential curves in the dictionary. A subset
of such a dictionary is shown in Fig.~\ref{linplot} (left). The other parameters are TR =
4.10 ms, 20 spokes per frame, 51 frames in total. The simulated curves are highly
correlated and can be represented by only a few principle components Fig.~\ref{linplot}. 
For easier comparison, the subspace-constraint
reconstruction used the coil sensitivity maps estimated using model-based
T1 reconstruction. The resulting linear problem was then solved using conjugate
gradient or FISTA algorithm in BART. The coefficient maps were then projected
back to image series where the 3‐parameter fit is applied for each voxel
according to Equation (\ref{signal_T1}). 

\begin{figure}
	\begin{center}
		\includegraphics[width=\textwidth]{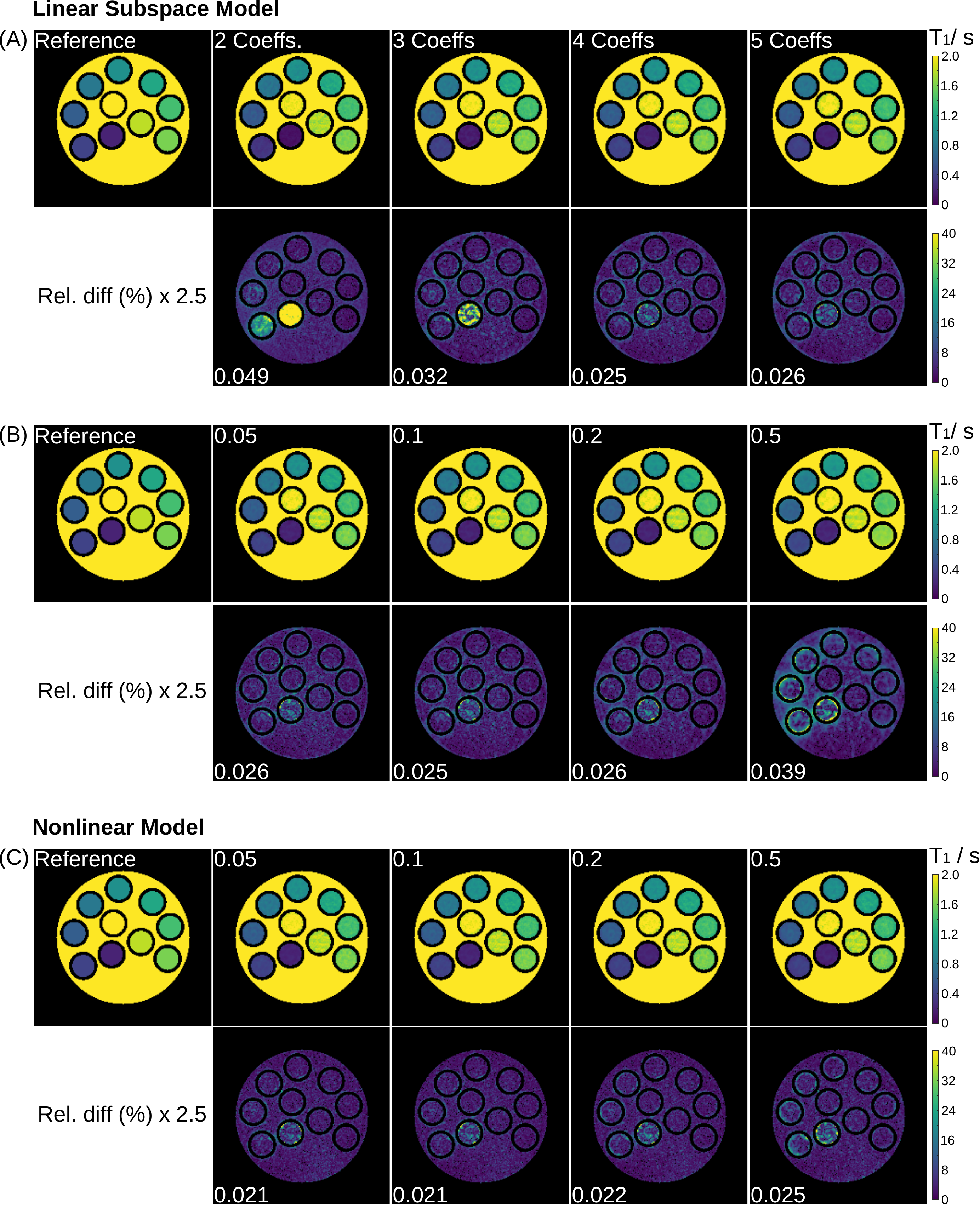}
	\end{center}
	\caption{Comparison of linear and nonlinear model-based reconstructions on the simulated phantom.
		\textbf{(A)}. Linear subspace reconstructed T1 maps using 2, 3, 4, 5 complex coefficients 
		and their relative difference to the reference. \textbf{(B)}. Linear subspace reconstructed 
		T1 maps using 4 complex coefficients with changing regularization parameters. 
		\textbf{(C)}. Model-based reconstructed T1 maps using different regularization strengths. 
		The numerical phantom used here is simulated using 208 frames, one spoke per frame, and TR = 20.5 ms. All reconstructions are done with L2-regularization. 
		The regularization strength and the normalized relative errors to the reference are shown on the top-left and bottom-left of each figure, respectively.}
	\label{linphantom}
\end{figure}

\begin{figure}
	\begin{center}
		\includegraphics[width=\textwidth]{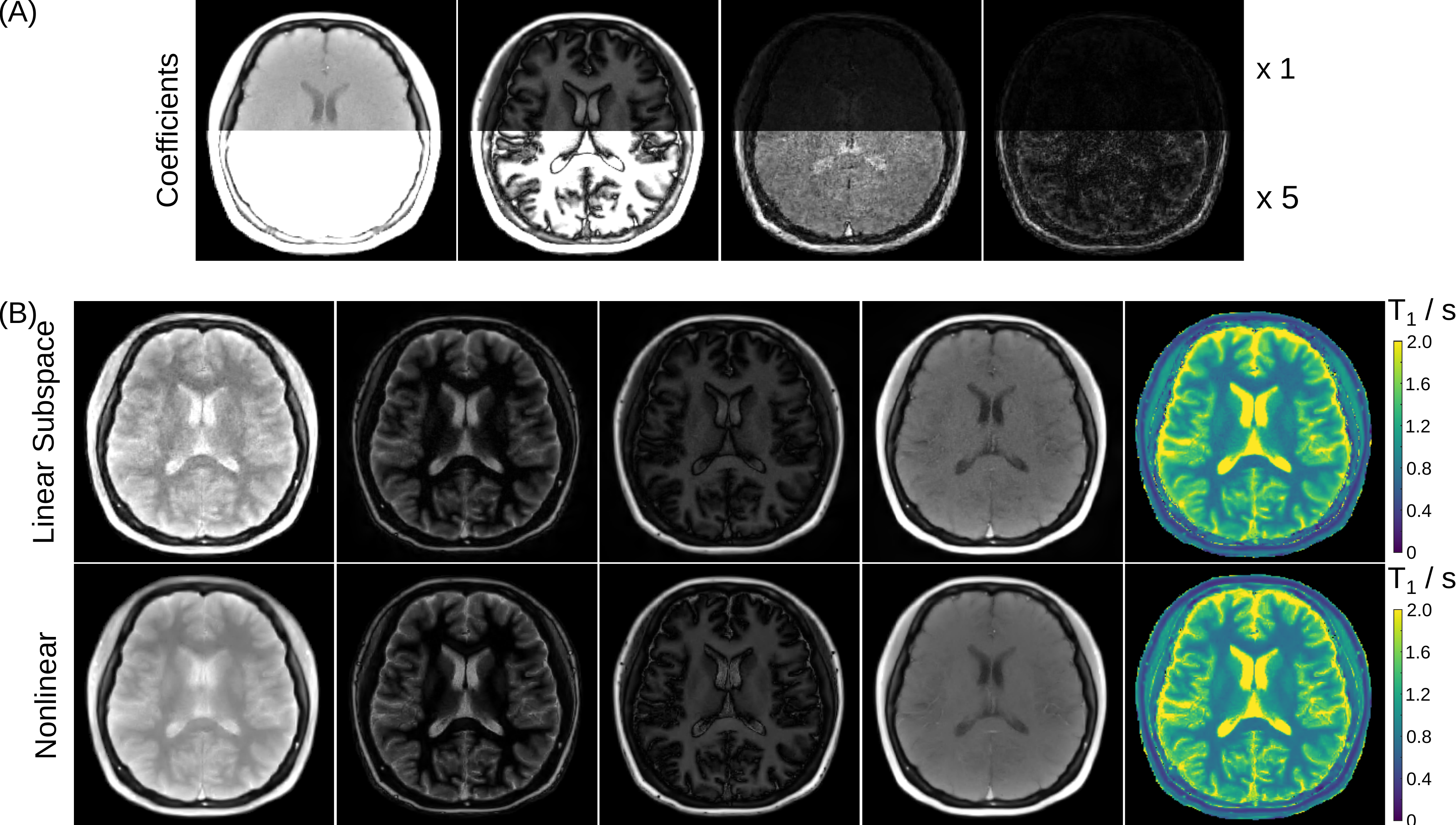}
	\end{center}
	\caption{\textbf{(A)} Reconstructed 4 complex coefficient maps (only magnitude is shown) using the 
		linear subspace method for a human brain study. 
		\textbf{(B)} Synthesized images (at inversion time 40 ms, 400 ms, 800 ms, 4000 ms) using (top)
		the above 4 complex coefficient maps of the linear subspace method and
		(bottom) the 3 physical maps of the nonlinear model-based reconstruction, respectively. The 
		corresponding T1 maps are presented in the rightmost column.}
	\label{linbrain}
\end{figure}

Fig.~\ref{linphantom} (A) shows estimated phantom T1 maps using a variant number of complex coefficients of the
linear subspace-based reconstruction with L2 regularization. Lower number of coefficients causes bias for
quantitative T1 mapping (especially for tubes with short T1s) while higher number of coefficients brings noise in the
final T1 maps. Therefore 4 coefficient maps were chosen to compromise between
quantitative accuracy and precision.  Fig.~\ref{linphantom} (B) compares the effects of regularization
strength. Similarly, low value of the regularization parameter brings noise while high regularization strength causes
bias. A value of $0.1$ was then chosen to compromise T1 accuracy and precision.
Fig.~\ref{linphantom} (C) then shows the effects of regularization for the model-based reconstruction.
A value of $0.1$ was selected as it has the least normalized error. 

The low normalized relative errors on the optimized T1 maps reflect both linear subspace and nonlinear model-based methods 
can generate T1 maps with good accuracy while nonlinear model-based reconstruction
has a slightly better performance (i.e., less normalized relative errors).

With the above settings, Fig.~\ref{linbrain} (A) depicts the four main coefficient maps estimated using the linear subspace
method for a brain study. In this case, a joint $\ell_{1}$-Wavelet sparsity regularization was applied to the maps with a strength of 
$0.0015$ to improve the precision. For this dataset, the reconstruction together with a pixel-wise fitting took around 2 minutes on the GPU.  Fig.~\ref{linbrain} (B) presents the synthesized images 
along with the corresponding T1 maps using (top) the 
above four coefficient maps for the linear subspace and (bottom) the 3 physical parameter maps for nonlinear model-based reconstructions, where
a similar joint $\ell_{1}$-Wavelet sparsity is applied with the regularization parameter $0.09$.
Again, both linear subspace and nonlinear methods could generate high-quality synthesized images and T1 maps while the nonlinear
methods have slightly less noise and better sharpness.

Although linear subspace reconstruction has been demonstrated to be a fast and
robust quantitative parameter mapping technique, it might not be directly
applicable to MR signals with phase modulation along echo trains. 
For instance, multi-gradient-echo signals are known to be modulated by off-resonance-induced
phases. A dictionary of multi-gradient-echo magnitude and phase signals was
simulated with $256 \times 256$ $T_2^*$ and $f_{B_0}$ combinations linearly ranging
from 1 to 100~\si{\ms} and from -200 to 200~\si{\hertz}, respectively.

Fig.~\ref{linplot} displays the magnitude
and phase evolution of 7 randomly-selected dictionary entries. The magnitude
signal follows the exponential decay, while phase wrappings occur with large
field inhomogeneity and long echo train readout. More importantly, the SVD
analysis of the signal dictionary shows that at least 26 principal components
are required to represent the complex signal behaviour.

\subsection{Frequency-Modulated SSFP}

\begin{figure}
	\begin{center}
		\includegraphics[width=\textwidth]{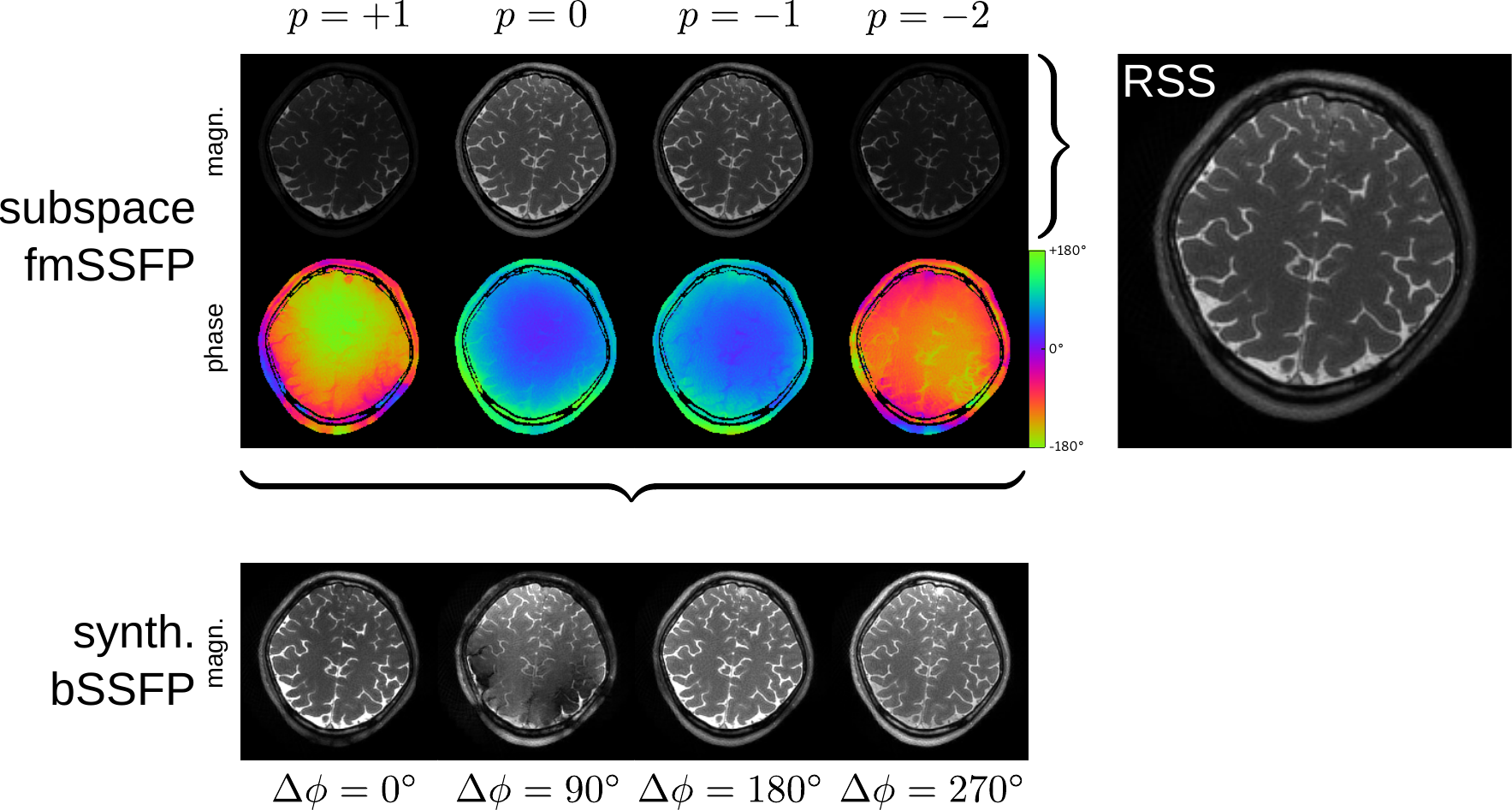}
	\end{center}
	\caption{
		Reconstructed subspace coefficients maps (top) along with
		its root-sum-squares composite image for a individual slice within the acquired
		3D volume. Synthesized bSSFP images are computed from these coefficient maps
		for different virtual frequency offsets (bottom).
	}
	\label{fig:fmSSFP}
\end{figure}

Conventional balanced steady-state free precession (bSSFP) sequences exhibit a
high signal-to-noise ratio (SNR) but suffer from possible signal voids in regions with certain off-resonance
distributions. These voids or banding artifacts can be removed when multiple images are 
acquired with different transmitter phase cycles. Foxall and coworkers demonstrated that
bSSFP sequences are tolerant to small but continuous changes in transmitter frequency
\cite{Foxall_Magn.Reson.Med._2002}. In \cite{Roeloffs_Magn.Reson.Med._2018} we exploited this
method to develop a time-efficient alternative to phase-cycled bSSFP that waives intermediate
preparation phases in phase-cycled bSSFP to establish different steady-states.
Image reconstruction is performed in the low-frequency Fourier subspace and yields signal intensity
and contrast comparable to on-resonant bSSFP.

To this end, a frequency-modulated SSFP (fmSSFP) pulse sequence \cite{Foxall_Magn.Reson.Med._2002}
was combined with 3D stack-of-stars data acquisition such that a single full
sweep through the spectral response profile was obtained.  Aligned partitions
allowed to decouple the reconstruction problem into individual slices by a 1D
inverse Fourier transform. After coil sensitivity estimation
\cite{Uecker_Magn.Reson.Med._2014}, image reconstruction was performed by
solving a linear subspace-constrained reconstruction problem using a local low
rank regularization.  As a subspace basis, the four
lowest order Fourier modes were chosen.  Figure 8 shows the reconstructed complex-valued coefficient maps from
which a composite image can be computed in a root-sum-squares manner
(top). Additionally, synthesized bSSFP images are computed for four
virtual frequency offsets (bottom) in which the distribution of signal
voids is given by the phase distribution of the subspace coefficients.
These synthesized bSSFP images correspond to conventional bSSFP images
acquired with four different phase cycles. 

\section{Discussion}

In the past decades various techniques were developed to accelerate quantitative MRI.
One very general way is to exploit complementary information from spatially distinct receiver coils, called parallel imaging (PI) \cite{
	Sodickson_Magn.Reson.Med._1997,
	 Pruessmann_Magn.Reson.Med._1999, 
	Griswold_Magn.Reson.Med._2002}.
Others make use of the fact that MR images are usually sparse in a certain transform domain and combined with incoherent sampling and nonlinear image reconstruction algorithms it is called compressed sensing (CS) \cite{Donoho_IEEETrans.Inform.Theory_2006}. Exploiting this prior knowledge about a compressible image, CS can recover MR images from highly undersampled data \cite{Lustig_Magn.Reson.Med._2007,Block_Magn.Reson.Med._2007}. Other approaches combine PI and CS with  efficient non-Cartesian sampling schemes \cite{Block_Magn.Reson.Med._2007}.\\
When it comes to parameter mapping, beside of the already mentioned sparsity constraints, also low-rank constraints or joint sparsity can be exploited along the parameter dimension to accelerate the acquisition time \cite{
	 Doneva_Magn.Reson.Med._2010,
	 velikina_Magn.Reson.Med.2013,
	 Zhang_Magn.Reson.Med._2015, 
	 Zhao_Magn.Reson.Med._2015,
	Maier_Magn.Reson.Med._2018}.\\
Generally speaking, the methods above usually 
consist of two steps: first reconstruction of contrast-weighted images from undersampled 
datasets and second, the subsequent voxel-by-voxel fitting/matching.
In contrast, model-based reconstructions 
integrate the underlying MR physics into the forward model, enabling estimation of MR physical images (parameter maps) directly 
from the undersampled $k$-space, bypassing the intermediate steps of image reconstruction and 
pixel-wise fitting/matching completely. This has the advantage of only reconstructing
the desired parameter maps instead of a set of contrast-weighted images, i.e., reducing 
the number of unknowns tremendously. Another advantage is that parameter
estimation using L2 norm in the data fidelity is optimal
(assuming white Gaussian noise), while fitting reconstructed magnitude
images may introduce a noise dependent bias.
Special sampling strategies are required for model-based
reconstruction to achieve good reconstructions. Sampling schemes 
used include CAIPIRINHA \cite{Breuer_Magn.Reson.Med._2005} and
golden-ratio radial acquisition \cite{Winkelmann_IEEETrans.Med.Imag._2007}.

In contrast to non-linear model-based reconstruction methods which use a minimal
number of physical parameters to describe the MR signal precisely,
linear subspace methods approximate the MR signal using a certain number
of principal coefficients. As discussed above, this computationally much more efficient
and avoids partial volume effects. Subspace methods were also successfully
used for multi-parametric imaging, for example using pseudo steady‐state
free precession (pSSFP)  \cite{Asslaender_Magn.Reson.Med._2018} or echo planar time-resolved
imaging where it is combined with non-linear iterative phase estimation
\cite{Dong_Magn.Reson.Med._2020}.
Subspace methods have to balance two
additional error terms coming from 1) the approximation error when the subspace
size is too small and 2) noise amplification when the subspace size is large 
(more unknowns). To minimize these additional errors, the optimal subspace size
has to be selected. While noise can be predicted based on the size of the
subspace, the approximation error is more difficult to control and may
require systematic studies that include comparisons to a ground truth.

Model-based reconstructions are, in general, memory demanding and time
consuming as all the data has to be hold in memory simultaneously during
iterations. However, modern computational devices such as GPUs have enabled
faster reconstructions. For example, the computation time for model-based T1
reconstruction presented here has been reduced from around 4 hours in CPU
(40‐core 2.3 GHz Intel Xeon E5‐2650 server with a RAM size of 512 GB) to 6
minutes using GPUs (Tesla V100 SXM2, NVIDIA, Santa Clara, CA). Other smart
computational strategies \cite{Mani_Magn.Reson.Med._2015,Uecker__2015,maier2020pyqmri}
may also be employed to reduce the memory and computational time. The other limitation might be that
model-based reconstructions are sensitive to model mismatch, e.g.,
bi-exponential processes, slow exchange regime. One way to overcome such
limitations is to explicitly model these effects and include them into the
model-based reconstruction.

Validation or the assessment of errors is an important part when developing
nonlinear model-based reconstruction methods. To this end, several strategies
should be applied. First, numerical simulations with analytical k-space models
can assure general convergence and robustness to noise as noise levels can be
freely chosen and noise-free ground truth is available. Second, in vitro or
phantom studies covering a certain range of parameters of interest should be
performed as effects such as intra-voxel dephasing, imperfect RF excitation, and
shimming are hard to simulate and the effect of such model errors are hard to
predict. Several hardware phantoms are commercially available, well
charactarized and widely used \cite{keenan2018}. Last, in vivo measurements
should always be evaluated against established methods or fully-sampled data
sets if possible.

Tremendous progress in the fields of machine learning / deep learning has
sparked a huge interest in applying these methods to different MRI applications
including image reconstruction \cite{Hammernik_Magn.Reson.Med._2017,
Aggarwal_IEEETrans.Med.Imag._2019}. However, so far only few applications exist
that target accelerated parameter mapping directly
\cite{Liu_Magn.Reson.Med._2019, Golkov_IEEETrans.Med.Imag._2016,
Cai_IEEETrans.Med.Imag_2019, Jun_ISMRM_2020}. While these are
promising developments, there are also still unsolved questions
regarding the stability of machine learning methods \cite{Antun_PNAS_2020}
and the risk of introducing image features that look real
but are not present in the data (hallucinations) \cite{muckley2020state}.

Magnetic Resonance Fingerprintig (MRF) \cite{Ma_Nature_2013} is an alternative
technique to perform time-efficient multi-parametric mapping leveraging high
undersampling factors. In its original formulation, parameter maps are
reconstructed in a two-step procedere. First, time series are generated by an
inverse NUFFT operation agnostic to any physical signal model. Second,
parameter maps are generated by pixel-wise matching of the obtained time series
with a precomputed dictionary consisting of simulated signal prototypes. The
proposed decoupling into a linear reconstruction of time series and a
nonlinear fitting problem solved by exhaustive search results in comparatively
short reconstruction times and does not require analytical signal models.
These two advantages rendered MRF a very popular approach in the recent years.
This two-step procedere, however, comes at a cost. The initial model-agnostic
gridding operation results in heavily aliased signal time courses. Aliasing
can be removed only partially by pixel-wise matching, as no information on the
sampling pattern is available in that step, and might deteriorate or bias the
obtained parameter maps. Recent works have tried to overcome this inherent
drawback of the two-step method by iterating between time and parameter domain
\cite{Davies_IEEEICASSP_2014} or by formulating the reconstruction as a
nonlinear problem that integrates the physical signal model and additional
image priors \cite{Zhao_IEEETrans.Med.Imag._2014} similar to the discussed
model-based approaches. Also techniques combining iterative reconstructions and
grid searches on dictionaries were developed
\cite{Zhao_IEEETrans.Med.Imag._2016}.  For a recent review that discusses the
basic concept of MRF also in the context of other quantitative methods see
\cite{Asslaender_Magn.Reson.Med._2020}.

\section{Conclusion}

By formulating image reconstruction as
an inverse problem, model-based reconstruction techniques can
estimate quantitative maps of the underlying physical
parameters directly from the acquired k-space signals
without intermediate image reconstruction. While this
is computationally demanding, it enables very efficient 
quantitative MRI.

\section*{Acknowledgments}

The authors would like to thank Mr. Ansgar Simon Adler for help with the phase-contrast flow MRI experiment,
Dr. Sebastian Rosenzweig for helpful dicussions and improvements to the figures, Dr. Tobias Block for
the radial spin-echo sequence, and Dr. Christian Holme for help with the scripts.

\section*{Funding Statement}

This work was supported by the DZHK (German Centre for Cardiovascular
Research), by the Deutsche Forschungsgemeinschaft (DFG, German Research
Foundation) under grant TA 1473/2-1 / UE 189/4-1 and under Germany’s Excellence
Strategy—EXC 2067/1‐390729940, and funded in part by NIH under grant
U24EB029240.

\section*{Data Accessibility}

All model-based reconstructions were performed with the BART toolbox.
Scripts to reproduce the examples shown in this work are
available at \url{https://github.com/mrirecon/physics-recon}. \\
Data is available at DOI:~10.5281/zenodo.4381986.

\section*{Competing Interests}

We have no competing interests.

\section*{Author's Contributions}

All authors contributed to the design of the studies, the acquisition
of the data, and the analysis. All authors helped to write the
manuscript and approved the final version.

\bibliographystyle{RS}
\bibliography{radiology.bib}

\end{document}